\documentclass[aip, apl,amsmath,amssymb,reprint,twocolumn]{revtex4-2}

\usepackage[utf8]{inputenc}
\usepackage{graphicx}
\usepackage{xcolor}
\usepackage{hyperref}
\usepackage{amsmath}

\usepackage[T1]{fontenc}
\usepackage{mathptmx}
\usepackage{etoolbox}
\usepackage{multirow}

\usepackage{array}
\usepackage{makecell}

\usepackage{booktabs}

\makeatletter
\def\@email#1#2{%
 \endgroup
 \patchcmd{\titleblock@produce}
  {\frontmatter@RRAPformat}
  {\frontmatter@RRAPformat{\produce@RRAP{*#1\href{mailto:#2}{#2}}}\frontmatter@RRAPformat}
  {}{}
}%
\makeatother

\begin{document}

\title{Perspective on traveling wave microwave parametric amplifiers}

\author{Martina Esposito}
\affiliation{Univ. Grenoble Alpes, CNRS, Grenoble INP, Institut N\'eel, 38000 Grenoble, France}
\author{Arpit Ranadive} 
\affiliation{Univ. Grenoble Alpes, CNRS, Grenoble INP, Institut N\'eel, 38000 Grenoble, France}
\author{Luca Planat}
\affiliation{Univ. Grenoble Alpes, CNRS, Grenoble INP, Institut N\'eel, 38000 Grenoble, France}
\author{Nicolas Roch$^*$}
\email[Corresponding author: ]{nicolas.roch@neel.cnrs.fr}
\affiliation{Univ. Grenoble Alpes, CNRS, Grenoble INP, Institut N\'eel, 38000 Grenoble, France}

\begin{abstract}	
	
		Quantum-limited microwave parametric amplifiers are genuine key pillars for rising quantum technologies and in general for applications that rely on the successful readout of weak microwave signals by adding only the minimum amount of noise allowed by quantum mechanics. In this perspective, after providing a brief overview on the different families of parametric microwave amplifiers, we focus on traveling wave parametric amplifiers (TWPAs), underlining the key achievements of the last years and the present open challenges. We discuss also possible new research directions beyond amplification such as exploring these devices as a platform for multi-mode entanglement generation and for the development of single photon detectors.   

	\end{abstract}



\maketitle 
\section{Introduction}

A low noise amplification chain is a key element of any system requiring the detection of low amplitude signals. In  microwave regime, a quantum-noise limited amplification chain is crucial, for example, for the readout of solid-state qubits, such as spin qubits~\cite{Stehlik2015} or superconducting qubits \cite{Krantz2019a}, but also in a broader range of modern applications such as nano-electromechanical systems~\cite{Teufel2011}, electron spin resonance detection~\cite{Bienfait2016}, astronomy instrumentation \cite{Smith2013,Bockstiegel2014} or axionic dark matter detection~\cite{Caldwell2017,Jeong2020,Backes2021a,Wurtz2021}.\\
The noise performance of an amplification chain is established primarily by the noise and gain characteristic of the first amplifier in the chain \cite{Pozar}. For this reason, it is highly desirable to achieve quantum-limited added noise and high gain at the first stage of amplification.\\
State-of-the-art commercially available semiconducting low-noise microwave amplifiers are unable to reach quantum-limited added noise because of their dissipative nature. Superconducting parametric amplifiers are instead well suited for this task since they allow to take advantage of highly non-linear interactions without introducing dissipation \cite{Devoret2013}. They can give access to quantum-limited noise performance with gain high enough to suppress the noise added by following amplifiers in a detection chain \cite{Clerk2010}.\\ 
The development of superconducting parametric amplifiers recently became topic of intense research efforts, given the great value of quantum-limited microwave detection chains in many modern research and application areas. \\
In this perspective, we briefly retrace the historic path that brought to the present state of the art for superconducting microwave parametric amplifiers, focusing in particular on traveling wave parametric amplifiers (TWPAs). For a more complete treatment of superconducting parametric amplification we redirect the reader to a recent review \cite{Aumentadoa}.\\
The perspective is divided in two sections. In Section \ref{historical_road-map}, we provide a brief historical road-map and a summary of the main approaches adopted for the development of TWPAs and of the obtained figures of merit. 
In Section \ref{open_challeges_and_future_durections}, we present issues that are currently still open in the development of such devices and discuss possible new research directions that could be of great relevance within and beyond the purpose of low noise amplification.

\section{\label{historical_road-map}A brief historical road-map}

Parametric amplification is a fundamental wave-mixing process in non-linear optics. When a material with non-linear polarization is excited/pumped with an intense electromagnetic field, a weak signal field can get amplified via the interaction with the medium. Energy gets transferred from the pump frequency mode $f_p$, to the signal frequency mode $f_s$ via the creation of a third mode $f_i$ dubbed idler \cite{Boyd}. Parametric amplification can occur either via a three wave-mixing process (3WM) where $f_s + f_i = f_p$, or via a four wave-mixing process (4WM) where $f_s + f_i = 2 f_p$. 

The signal gain can be enhanced by increasing the interaction time in the non-linear medium via two main strategies. 
The first one,  resonant-amplification, consists in placing the nonlinear medium in a cavity; in this way the interaction time will be as long as the inverse of the cavity line-width. This approach however puts a constraint on the amplification bandwidth which depends on the cavity line-width as well. The second strategy, traveling wave amplification, consists in optimizing the gain by increasing the physical length of the non-linear medium, removing the constraint on the amplification bandwidth given by the presence of a cavity.  

For microwave parametric amplification, the role of the non-linear medium can be played by the non-linear inductance of a superconducting circuit composed for example of Josephson junctions. 
Josephson junction based amplification reaching near quantum limited operation was demonstrated for the first time by Bernard Yurke's pioneering works in the late 1980's \cite{Yurke1988,Yurke1989}. 
In the early 2000's, with the arising of the field of circuit quantum electrodynamics (c-QED) \cite{Vion2002a,Blais2004,Wallraff2004,Koch2007}, a renovated strong interest in microwave quantum limited amplification started and continued for the last two decades with intense research activity in many groups.

The fervent effort in developing such devices was strongly motivated by the pursuit of obtaining single-shot and high fidelity qubit readout in c-QED. This breakthrough result was initially achieved with the Josephson bifurcation amplifier (JBA)~\cite{Siddiqi2004,Manucharyan2007,Mallet2009}, that exploited the bistable regime of an rf-driven Josephson junction.   
	\begin{figure*}[thb]
		\centering
		\includegraphics[width=0.9\textwidth]{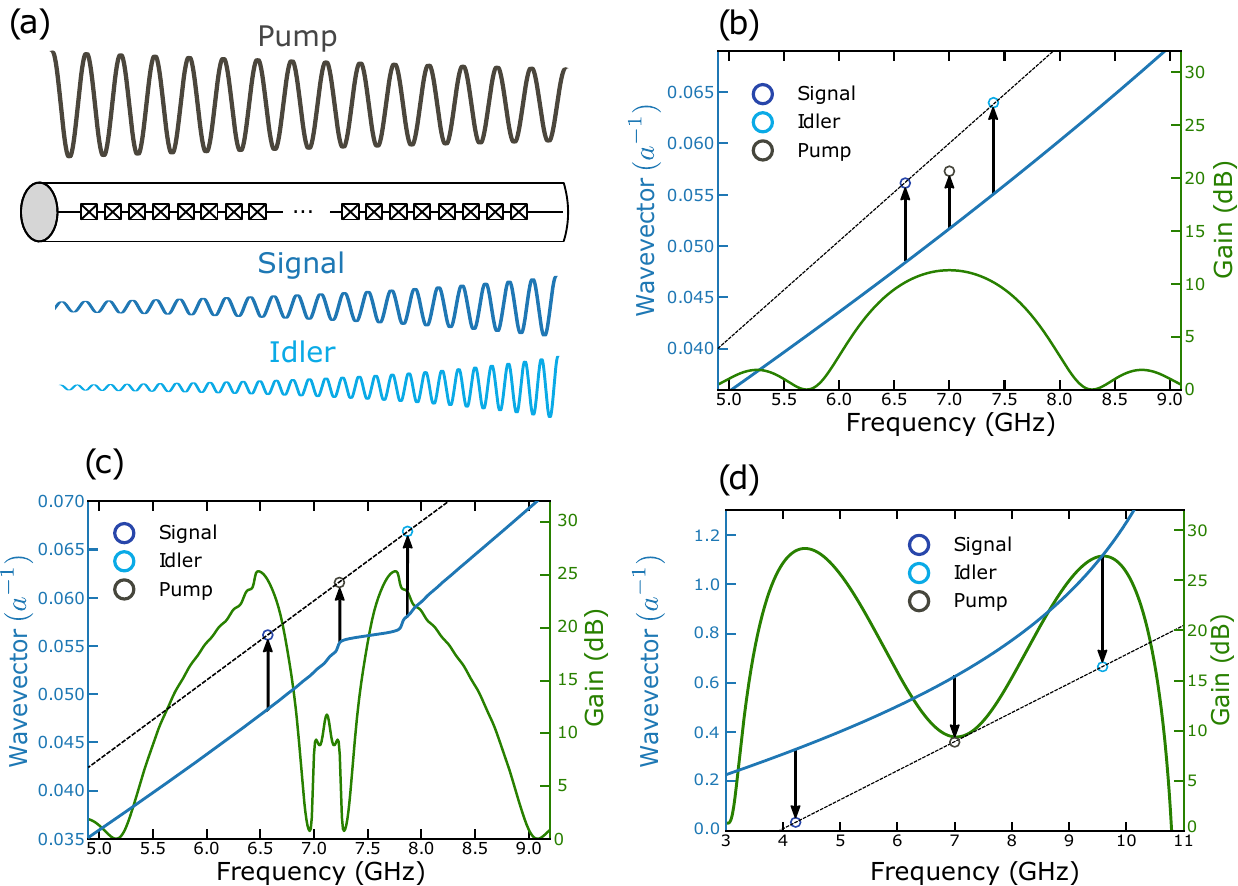}
		\caption{\label{Fig1} \textbf{Cartoon of TWPA physics and phase matching approaches.} (a) Cartoon of a TWPA composed by a chain of Josephson junctions with propagating pump, signal and idler fields. Simulated gain (green curve) and chromatic dispersion relation (blue curve) for (b) a bare TWPA (no-phase matching approaches adopted), (c) a TWPA with dispersion engineering phase matching and (d) a TWPA with reversed Kerr phase matching. The wave-vector is in units of $1/a$, where $a$ is the size of a unit cell. The arrows indicate the Kerr phase shift. The circles indicate the wave-vectors for pump, signal and idler. Phase matching is reached when the three wave-vectors lay on the same line. Dashed lines are guides for the eye. Dispersion relation curves and Kerr phase shifts are exaggerated for pedagogical purposes. Gain curves are simulated using standard coupled modes equations and typical TWPAs parameters \cite{Planat2020b,Ranadive2021}. The code used for simulations of gain profiles and dispersion relations is available at \cite{link_code}. }
	\end{figure*}
Subsequently, Castellanos-Beltran \textit{et. al} successfully implemented what then became an established circuit scheme for resonant Josephson parametric amplifiers, resonant-JPAs~\cite{Castellanos-Beltran2007}{}.
Since then, resonant-JPAs reaching the quantum limit of noise have been successfully demonstrated and operated in many labs. Remarkable progresses have been obtained regarding their bandwidth~\cite{Mutus2014,Roy2015}, their saturation point~\cite{Castellanos-Beltran2008,Liu2017,Planat2019c}, pumping schemes~\cite{Yamamoto2008,Bergeal2010a,Roch2012,Eichler2014b,Frattini2018a}, frequency tunability \cite{Sivak2020} and non-reciprocity~\cite{Abdo2011,Metelmann2015,Lecocq2017,Lecocq2020}.\\
Despite such notable improvements, the bandwidth constraint imposed by the presence of the cavity in the circuit scheme of resonant-JPAs still represents a significant limitation for some applications, such as multiplexed readout of high number of qubits or broadband squeezing generation. This strongly motivates the development of TWPA devices, superconducting non-linear meta-materials that exhibit parametric amplification in a traveling wave configuration. TWPAs allow to achieve, with superconducting circuits, what can be achieved at optical frequencies with nonlinear optical fibers \cite{Agrawal1990}, with the advantage to operate over a much larger fractional bandwidth.\\ 
%

TWPA devices \cite{CULLEN1958, Tien58, cullen60, Sorenssen1962} are more challenging to develop in comparison to their cousins resonant-JPAs. The necessity to engineer a long non-linear medium requires typically more complex nano-fabrication methods. In addition, efficient traveling wave amplification requires the fulfillment of two challenging conditions: the impedance matching of the amplifier with the typical $50 \, \Omega$ impedance of the external environment, and the attainment of the phase matching condition for the pump, signal and idler fields. \\
Early theoretical and experimental TWPA investigations emerged already in the 70s-90s \cite{Feldman75,Wahlsten77, Sweeny85, Yurke96}. However, it was in 2010s that TWPAs started to be conceived as we know them today and successfully realized with modern nano-fabrication techniques \cite{Slichter2010,Slichter2011_thesis}.         
%
Key progresses were obtained with the demonstration of TWPA devices exploiting the high kinetic inductance of disordered superconductors~\cite{HoEom2012b}, or Josephson junctions~\cite{White2015}, and in 2015 the first demonstration of near quantum-noise-limited TWPA device was achieved \cite{Macklin2015}.\\  
These breakthrough achievements demonstrated the possibility to go beyond the bandwidth constraints of resonant-JPAs and opened the road for the exploration of TWPAs based both on kinetic inductance of superconducting thin films (KTWPAs) and on Josephson junction based meta-materials (JTWPAs). \\
In Fig \ref{Fig1} (a) we show a sketch on a TWPA consisting in a chain of Josephson junctions. For the case of KTWPAs, the array of Josephson junctions is substituted with a long high kinetic inductance superconducting transmission line.\\

\noindent \textbf{Challenges and adopted solutions}\\
Several strategies have been adopted for addressing the issues of complex nano-fabrication, impedance matching and phase matching.
Among all, phase matching (momentum conservation) is perhaps the biggest challenge in TWPAs development. 
Phase mismatch between signal idler and pump fields in TWPA devices arises because of two reasons: (1) dispersion phase mismatch, caused by the curved chromatic dispersion relation \cite{Zorin2016} of the meta-material; (2) Kerr phase mismatch, caused by non-linear processes, such as self phase modulation (SPM) and cross phase modulation (XPM) \cite{Agrawal1990}, making the pump traveling faster (with higher phase velocity) than the signal and the idler. Dispersion and Kerr phase mismatch typically sum up giving a total phase mismatch that limits the optimal gain. \\
The first strategy, dispersion engineering phase matching, that has been adopted to solve such problem is based on engineering a gap in the dispersion relation of the meta-material. Such modification of the dispersion relation allows to fulfill the phase-matching condition at a specific pump frequency close to the gap. Dispersion engineering phase matching has been obtained with different experimental approaches: using periodic loading patterns \cite{HoEom2012b,Planat2020b} and via the introduction of resonant elements alongside the meta-material (resonant phase matching) \cite{OBrien2014}.\\
Recently, an alternative to dispersion engineering phase matching based on the sign reversal of the Kerr nonlinearity \cite{Bell2015} has been experimentally demonstrated \cite{Ranadive2021}, reversed Kerr phase matching. This phase matching mechanism is based on reversing the sign of the Kerr phase mismatch such that it compensates the dispersion phase mismatch, cancelling the total one. The sign reversal of the Kerr phase mismatch is obtained by using a superconducting nonlinear asymmetric element (SNAIL) \cite{Frattini2017b} as units cell of the TWPA, allowing in-situ tunability of the 3WM and 4WM nonlinearities via an external magnetic flux. 
In a reversed Kerr TWPA, optimal gain can be reached for a large range of pump frequencies.\\ 
In Fig \ref{Fig1} (b)-(d)  we elucidate the physics of phase matching mechanisms in TWPAs with pictorial sketches. In panel (b) we shown gain and chromatic dispersion relations for a bare TWPA with no phase matching approach, giving a low gain limited by the total phase mismatch. The arrows indicate the Kerr phase mismatch. In panel (c), we show the same for a TWPA with dispersion engineering phase matching and in panel (d) reversed Kerr phase matching.  \\
In Table \ref{table1}, different experimental approaches adopted so far for the phase matching are summarized, together with the techniques implemented for addressing the other, not less difficult, challenges of nano-fabrication and impedance matching.
\begin{table*}
\caption{\label{table1} Summary of TWPA devices and respective approaches used for addressing the challenges of nano-fabrication, impedance matching and phase matching.}
\begin{ruledtabular}
\begin{tabular}{l|lll}
  &   Nano-fab & Impedance matching & Phase matching \\ \hline

2012 Ho Eom et al.\cite{HoEom2012b}& \multirow{4}{*}{NbTiN CPW}& \multirow{4}{*}{Tapered impedance transformers}& \multirow{4}{*}{Periodic loading pattern}\\  
 2014 Bockstiegel et al. \cite{Bockstiegel2014}&&&\\
 2016 Vissers et al.\cite{Vissers2016}&&&\\
 2018 Ranzani et al. \cite{Ranzani2018}&&&\\
 \hline
 2015 Macklin et al.\cite{Macklin2015}& \makecell[l]{ Nb/Al-AlO$_x$/Nb JJ-array \\ Nb trilayer process}& Lumped-element capacitors& Resonant Phase Matching (RPM) \cite{OBrien2014}\\ 
 \hline
 2015 White et al. \cite{White2015}& \makecell[l]{ Al/Al$_{2}$O$_3$/Al  JJ-array \\ Double angle evaporation} & Lumped-element capacitors& Resonant Phase Matching (RPM) \cite{OBrien2014}\\
 \hline
2016 Adamyan et al. \cite{Adamyan2016}& NbN CPW & Quasi-fractal interdigitated capacitors& Periodic loading pattern\\
 \hline
 2017 Chaudhuri et al. \cite{Chaudhuri2017a}& \makecell[l]{ NbTiN  Lumped-elements}& Fish-bone interdigitated capacitors& Periodic loading \& Resonance Phase Shift\\
 \hline
 2018 Zorin et al. \cite{Zorin2018}& \multirow{2}{*} {\makecell[l]{Nb/AlO$_x$/Nb JJ-array \\ Nb multilayer process}}&\multirow{2}{*} {Lumped-element capacitors}&\multirow{2}{*}{---}\\ 
 2019 Miano net al. \cite{Miano2019}& & &\\
 \hline
 2019 Planat et al.\cite{Planat2020b}& \makecell[l]{ Al/Al$_{2}$O$_3$/Al  JJ-array \\ Double angle evaporation}& Top ground plane, microstrip geometry& Periodic JJs modulation\\ 
 \hline
 2020 Goldstein et al. \cite{Goldstein2020}& WSi microstrip& Top ground plane, microstrip geometry& Periodic loading pattern\\
 \hline
 2021 Malnou et al.\cite{Malnou2021}& NbTiN CPW& Fish-bone interdigitated capacitors& Periodic loading pattern \\
 \hline
 2021 Ranadive et al.\cite{Ranadive2021}& \makecell[l]{ Al/Al$_{2}$O$_3$/Al  JJ-array \\ Double angle evaporation}& Top ground plane, microstrip geometry& Reversed Kerr phase matching \cite{Bell2015}\\
  \hline
 2021 Shu et al.\cite{Shu2021}& NbTiN microstrip& Top ground plane, microstrip geometry& Periodic loading pattern\\
\end{tabular}
\end{ruledtabular}
\end{table*}
%
TWPAs' performances are typically evaluated on the basis of some key figures of merit: gain, bandwidth, added noise, saturation power and required pump power. The typical gain target is of the order of 20 dB. Such value guarantees that, when the TWPA is placed at the beginning of a standard microwave detection chain, the system noise of the entire chain is mostly dominated by the added noise of the TWPA \cite{Pozar}. The total system noise target is the standard quantum limit (SQL). The bandwidth and the saturation power are typically targeted to be as large as possible to allow the broadest range of possible applications. Finally, the required pump power should ideally be as low as possible to avoid saturation of the next stages of amplification.\\
To give the reader a feeling of the last progresses in improving such figures of merit, in Table \ref{tab:table2} we report a comparison for different TWPA devices. 
\begin{table*}
			\caption{\label{tab:table2} Comparison of TWPA figures of merit. For each reported TWPA, we list the kind of wave-mixing process, total system noise of the amplification chain having the TWPA as first stage of amplification (SQL corresponds to 1 photon total system noise: half photon plus additional half photon added by the amplified idler vacuum), best gain value, bandwidth (defined considering the frequency region with gain within 3dB from the maximum gain), saturation power (-1dB compression point) and pump power necessary to reach the maximum gain.}
			\begin{ruledtabular}
			\begin{tabular}{l| c c c c c c} 
					& Process & System noise [photons] & Gain [dB] & Bandwidth [GHz] & Saturation [dBm] & Pump power [dBm]\\ 
				\hline
				2012 Ho Eom et al.\cite{HoEom2012b} &  4WM & --- & 10 & 4.5  & -52 & -9 \\
				\hline
				2014 Bockstiegel et al. \cite{Bockstiegel2014} &  4WM & --- & 20 & 7  & --- & -10 \\
				\hline
				2016 Vissers et al.\cite{Vissers2016} &  3WM & --- & 18 & 4  & -35$^*$ & -15$^*$ \\
				\hline
				2018 Ranzani et al. \cite{Ranzani2018} &  3WM & 3.5 & 10 & 5  & -40 & -30 \\
				\hline
				2015 Macklin et al.\cite{Macklin2015} &  4WM & 2 & 21.6 & 3  & -99 & -63 \\
				\hline
				2015 White et al. \cite{White2015} &  4WM & 4 & 12 & 4  & -90 & --- \\
				\hline
				2016 Adamyan et al. \cite{Adamyan2016}& 4WM & --- & 6 & ---  & --- & -15 \\
				\hline
				2017 Chaudhuri et al. \cite{Chaudhuri2017a} &  4WM & --- & 15 & 3  & --- & --- \\
				\hline
				2018 Zorin et al. \cite{Zorin2018} &  3WM & --- & 11 & 3  & --- & --- \\
				\hline
				2019 Miano net al. \cite{Miano2019} &  3WM & --- & 12.5 & 4  & --- & --- \\
				\hline
				2019 Planat et al.\cite{Planat2020b} & 4WM & 3 & 18 & 2.3  & -100 & -70 \\
				\hline
				2020 Goldstein et al. \cite{Goldstein2020} & 4WM & --- & 15 & 2  & --- & -53 \\
				\hline
				2021 Malnou et al.\cite{Malnou2021}& 3WM & 3 & 16.5 & 2  & -63 & -29 \\
				\hline
				2021 Ranadive et al.\cite{Ranadive2021} & 4WM & 4 & 20 & 3  & -97 & -75 \\
                 \hline
                2021 Shu et al.\cite{Shu2021}& 3WM & --- & 18 & 19 & -57 & -29 \\
			\end{tabular}
			\end{ruledtabular}
		\begin{flushleft}
		When not explicitly stated, gain/bandwidth values are approximated from the published gain plots. \\
		$^*$ Measured for 10 dB gain.
		\end{flushleft}
	\end{table*}

\section{\label{open_challeges_and_future_durections}Open challenges and outlook}
Despite the remarkable improvements of the last few years in TWPAs development, several challenges are still open. In the following, we list what we think are are the most relevant open issues, discussing possible approaches to address them and potential future research directions.\\

\noindent \textbf{Reaching standard quantum limit (SQL)}\\
Improving added noise in TWPA is certainly one of the major open challenges. So far, the system noise of microwave detection chains which employ a TWPA as first stage of amplification has been reported to be at best two times larger than SQL (Table \ref{tab:table2}). Although full understanding is still missing, there are two main candidate phenomena that could explain it: capacitive dielectric losses and sidebands generation.\\
Dielectric losses are coming from the dielectric in the capacitive elements that are typically inserted in the nonlinear meta-materials to reach impedance of 50 $\Omega$. Enhancing the quality of such dielectric is one of the main strategies to reduce losses. Dielectrics with low number of defects such as amorphous silicon\cite{Oconnell2008} or silicon nitride \cite{Paik2010a} reported loss tangent between $10^{-6}$ and $10^{-5}$ and are promising candidates for reducing losses and possibly improve added noise in TWPAs. Another interesting approach would be the complete removal of the dielectric material and the use of vacuum gap capacitors\cite{Boussaha2020a}.\\
Further source of non-ideal added noise in TWPAs is the generation of sidebands, given by the intrinsic broadband/multi-mode nature of TWPA devices. A recently suggested strategy to mitigate the effect of sidebands generation on the added noise is based on a Floquet Mode TWPA implementation\cite{Peng2021}. Sidebands could also be mitigated via dispersion engineering approaches.\\

\noindent \textbf{Saturation power improvement} \\
Saturation power defines the maximum signal input power. Increasing the signal input power handling capability of TWPAs is a key goal especially for the commercial adaptability of these devices. The 1dB compression point (signal power at which the gain is reduced of 1 dB) in state-of-the-art TWPAs is still considerably inferior with respect to commercially available semiconductor-based amplifiers. A possible source of this limitation is pump depletion, coming from the inevitable conversion of pump power into power at signal and idler frequency or from possible leaks of power to higher harmonics \cite{Zorin2016}. This explains why KTWPAs, that work at pump powers much higher than JTWPAs, presents higher saturation powers. However, the necessity of high pump powers (typically 4-5 orders of magnitude higher than JTWPAs) have the disadvantage of requiring pump cancellation methods to avoid saturation of the next stages of amplification. Another possible sources of limited saturation power could be Kerr phase modulation and higher order processes \cite{Liu2017,Dixon2020}.\\

\pagebreak
\noindent \textbf{Pumping schemes}\\
Wave mixing mechanisms, 3WM and 4WM, determine the pumping scheme of a TWPA. For 4WM amplification, the pump frequency lies in the middle of the amplification band, while, for 3WM amplification, pump frequency is far away from the amplification range since it is set to be equal to the sum of signal and idler frequencies. KTWPAs have been demonstrated experimentally both in the 3WM and in the 4WM configuration, while JTWPAs so far only in the 4WM configuration. The advantage of 3WM amplification is the absence of the pump frequency in the amplification band. However, one needs to adopt strategies to mitigate unwanted 3WM processes like second harmonic generation, that is typically more efficient than parametric amplification.\\ 
Another advantage of 3WM amplification is the possibility to be implemented in theory with a flux-pumping scheme. TWPAs experimentally demonstrated so far are pumped by a microwave tone which travels through the meta-material together with the signal. One limitation of this current-pumping scheme is that it is very difficult to separate the pump from the signal field, causing potential detrimental effects to the system under measurement. A flux-pumping approach\cite{Zorin2019} instead would have the advantage of physically separating the pump input line from the signal input line providing better isolation. \\

\noindent \textbf{Non reciprocity}\\
TWPA devices are in principle directional since the signal gets amplified in the direction of the pump propagation. However, they can still transmit in the direction opposite to gain  
and a careful analysis of wave propagation in non-linear materials shows that under certain conditions modes with opposite wave-vectors can also be coupled leading to backwards gain\cite{Planat2020b,Peng2021}. This means that vacuum fluctuations could be backward amplified and exposed to the device under test. Hence, even if technical problems such as pump leakage and impedance matching could be solved with better engineering, isolators between the device under measurement, for example a qubit, and the TWPA would still be required.\\  
The demonstration of a truly directional TWPAs, with signal attenuation in the direction opposite to gain, would be a remarkable advancement in the field.\\ 
Directional amplification in narrow-band resonant-JPAs has been successfully achieved via reservoir engineering and multi-mode parametric interaction \cite{Metelmann2015,Sliwa2015,Lecocq2017}. 
Directional isolation in Josephson transmission lines has been demonstrated by frequency conversion\cite{Ranzani2017a} and recently adiabatic mode conversion has been also proposed \cite{Naghiloo2021}. Combining traveling wave parametric amplification and broadband backward isolation would be of great impact for superconducting quantum technologies.\\

\noindent \textbf{Operating frequency range and operating temperature}\\
TWPA devices developed so far are usually operating in the typical c-QED microwave frequency range, between 4 GHz to 12 GHz. The operation of TWPAs in other frequency ranges recently became subject of intense research. For example, TWPAs working at frequencies lower than 1 GHz are are very appealing for the spin qubits community~\cite{Stehlik2015}, while higher frequencies are required for some applications such as the search of dark-matter axions \cite{Caldwell2017}.
In the context of astronomy instrumentation, a valuable achievement would be also the possibility to operate TWPAs at $4$ K temperature, in order to substitute the dissipative semiconductor amplifiers in large detector arrays \cite{Bandler2019} with superconducting amplifiers, drastically reducing the power requirement for amplification. In this case, it will be necessary to employ materials that are still superconductors at $4$ K, for example NbTiN or also high $T_C$ superconductors.\\

\noindent \textbf{Beyond amplification}\\
As we reported in this paper, superconducting non-linear meta-materials became an established solution to reach near quantum-noise-limited broadband microwave amplification and serve as optimal pre-amplifiers in many context. Nonetheless, they have the potential to be successfully adopted for implementations that go also beyond quantum-limited amplification.   
They could serve as a powerful platform for multi-mode entanglement generation \cite{Grimsmo2017,Fasolo2021} and potentially quantum information with continuous variables. Their broadband nature and the possibility of in-situ customization of the nonlinearities\cite{Ranadive2021} are unique features for quantum optics investigations in the microwave regime. In addition, the intrinsic multi-mode nature of TWPA devices, that is considered detrimental for added noise and saturation power of an amplifier, could became instead a resource for the generation of multi-mode entanglement.\\
Moreover, inspired by JTWPAs, single-photon counters have been recently proposed \cite{Grimsmo2021}. By utilizing the tunability of strength and nature of non-linearity of Josephson meta-materials, these detectors could allow broadband and non-destructive counting of traveling microwave photons, opening new possibilities for single-photon measurement and control in c-QED.

\section{Conclusions}
In this perspective we reviewed progresses in superconducting traveling wave parametric amplifiers. We presented the main obstacles to the development of such devices and the different approaches adopted so far to overcome them, reporting and comparing the corresponding figures of merit. We discussed open issues related to the achievement of the standard quantum limit, the improvement of saturation power, the adoption of different pumping schemes, the implementation of non reciprocity and the extension to different frequency ranges and higher working temperatures. Future progresses in all these areas would be of great value for a diverse range of applications requiring broadband and low-noise microwave amplification, from quantum computation with solid state qubits to astronomy detection. Finally, we reported examples in which superconducting nonlinear meta-materials could be extremely valuable for purposes also beyond amplification such as multi-mode entanglement generation and detection of single traveling photons, appealing tools for quantum optics and metrology in the microwave frequency domain.

\section*{Acknowledgments}
This work is supported by the European Union’s Horizon 2020 research and innovation program under grant agreement no. 899561.  M.E. acknowledges the European Union’s Horizon 2020 research and innovation program under the Marie Sklodowska Curie (grant agreement no. MSCA-IF-835791). A.R. acknowledges the European Union’s Horizon 2020 research and innovation program under the Marie Sklodowska Curie grant agreement No 754303 and the ’Investissements d’avenir’ (ANR-15-IDEX-02) programs of the French National Research Agency.
The authors are grateful to the members of the superconducting circuits group at Neel Institute for helpful discussions.
\section*{Data Availability}
Data sharing is not applicable to this article as no new data were created or analyzed in this study.

\section*{REFERENCES}

%

\end{document}